\documentclass[10pt,conference]{IEEEtran}
\usepackage{mathptmx}

\usepackage{cite}

\ifCLASSINFOpdf
\else
\fi
\usepackage{hyperref}
\usepackage{soul}
\usepackage{rotating}
\usepackage{longtable}
\usepackage{booktabs}

\usepackage{pdfpages}

\usepackage{wrapfig}
\usepackage{amsmath}
\usepackage{graphicx}
\usepackage{caption}
\usepackage{footnote}

\usepackage{listings}

\definecolor{violet}{cmyk}{0.45,0.97,0.27,0.21}
\definecolor{lstblue}{cmyk}{1,0.80,0,0}
\definecolor{lstgreen}{cmyk}{0.71,0.21,0.65,0.22}
\definecolor{bluegrey}{cmyk}{0.56,0.24,0.11,0.05}
\definecolor{javadoc}{cmyk}{0.88,0.59,0,0}
\definecolor{lstgrey}{cmyk}{0.55,0.44,0.42,0.32}

\lstdefinelanguage{SQL}{
     keywords={},
     keywordstyle=\color{bluegrey}\bfseries,
     morekeywords=[2]{CREATE,TABLE,IF,NOT,EXISTS,NULL,SET,DEFAULT,PRIMARY,KEY,COLLATE,CHARACTER,AUTO_INCREMENT,ENGINE,CHARSET},
     keywordstyle={[2]\color{violet}\bfseries},
     otherkeywords={int,varchar,double,text,tinyint},
     sensitive=false,
     morecomment=[l][\color{lstgreen}]{//},
     morecomment=[s][\color{lstgreen}]{/*}{*/},
     morecomment=[s][\color{javadoc}]{/**}{*/},
     morestring=[b]',
     morestring=[b]"
  }
\lstdefinelanguage{PHP}{
     keywords={},
     keywordstyle=\color{bluegrey}\bfseries,
     morekeywords=[2]{static,function,if,return,pow,sin,cos,asin,min,sqrt,int},
     keywordstyle={[2]\color{violet}\bfseries},
     otherkeywords={@param, @returns, @author, @type, @link, @see},
     sensitive=false,
     morecomment=[l][\color{lstgreen}]{//},
     morecomment=[s][\color{lstgreen}]{/*}{*/},
     morecomment=[s][\color{javadoc}]{/**}{*/},
     morestring=[b]',
     morestring=[b]"
  }
\lstdefinelanguage{JavaScript}{
     keywords={},
     keywordstyle=\color{bluegrey}\bfseries,
     morekeywords=[2]{attributes, class, classend, do, empty, endif, endwhile, fail, function, functionend, if, implements, in, inherit, inout, not, of, operations, out, return, set, then, types, while, use},
     keywordstyle={[2]\color{violet}\bfseries},
     otherkeywords={@param, @returns, @author, @type, @link, @see},
     sensitive=false,
     morecomment=[l][\color{lstgreen}]{//},
     morecomment=[s][\color{lstgreen}]{/*}{*/},
     morecomment=[s][\color{javadoc}]{/**}{*/},
     morestring=[b]',
     morestring=[b]"
  }
\lstdefinelanguage{Java}{
     basicstyle=\ttfamily\scriptsize,
     keywords={},
     keywordstyle=\color{bluegrey}\bfseries,
     morekeywords=[2]{abstract,boolean,break,byte,case,catch,char,class,
      const,continue,default,do,double,else,extends,false,final,
      finally,float,for,goto,if,implements,import,instanceof,int,
      interface,label,long,native,new,null,package,private,protected,
      public,return,short,static,super,switch,synchronized,this,throw,
      throws,transient,true,try,void,volatile,while},
     keywordstyle={[2]\color{violet}\bfseries},
     morekeywords=[3]{@SuppressWarnings, @Capability, @Override},
     keywordstyle={[3]\color{lstgrey}},
     otherkeywords={@param, @return, @returns, @author, @link, @see},
     sensitive,
     morecomment=[l]//,
     morecomment=[s]{/*}{*/},
     morecomment=[s][\color{javadoc}]{/**}{*/},
     morestring=[b]",
     morestring=[b]',
  }[keywords,comments,strings]

\begin{document}
%
\title{Cloud Service Matchmaking using Constraint Programming}

\author{\IEEEauthorblockN{Beg{\"u}m {\.I}lke Zilci, Mathias Slawik, and Axel K{\"u}pper}
\IEEEauthorblockA{Service-centric Networking\\Technische Universit{\"a}t Berlin, Germany \\
\{\textit{ilke.zilci}\textbar \textit{mathias.slawik}\textbar \textit{axel.kuepper}\}@\textit{tu-berlin.de}}
}


%


\maketitle
\thispagestyle{plain}
\pagestyle{plain}

\begin{abstract}
Service requesters with limited technical knowledge should be able to compare services based on their quality of service (QoS) requirements in cloud service marketplaces. Existing service matching approaches focus on QoS requirements as discrete numeric values and intervals. The analysis of existing research on non-functional properties reveals two improvement opportunities: list-typed QoS properties as well as explicit handling of preferences for lower or higher property values. We develop a concept and constraint models for a service 
matcher which contributes to existing approaches by addressing these issues using constraint solvers. The prototype uses an API at the standardisation stage and discovers implementation challenges. This paper concludes that constraint solvers provide a valuable tool to solve the service matching problem with soft constraints and are capable of covering all QoS property types in our analysis. Our approach is to be further investigated in the application context of cloud federations.
\end{abstract}


%
\IEEEpeerreviewmaketitle

\section{Introduction}

Service brokering is a vivid research area where recent publications with several
approaches can be found targeting various contexts such as cloud service marketplaces and cloud federations.

\textit{Cloud service marketplaces} are platforms which act as a mediator between the service requesters and service providers. These platforms discover and store services in a service repository and allow service requesters to browse, select and interact with services via a \textit{service broker} component. They provide a unified view of the cloud service descriptions to the service requester and in some cases additional functionality such as unified monitoring, billing, and enhanced single sign on services \cite{appexchange}, \cite{googlemarketplace}, \cite{awsMarketplace}, \cite{thatmann2012towards}.
 
\textit{Cloud federations} are defined as inter-cloud organisations which comprise a set of autonomous and heterogeneous clouds \cite{grozev2014inter}. Initial discussions on cloud federations bring up the question on how the service requester can keep control of the selected clouds. For this, it is suggested that the service requester should be able to set requirements to be fulfilled while the federation distributes the deployment on several clouds. A cloud federation should "respect end-to-end SLAs" (service level agreements) \cite{cloudFed2014}. To ensure this, a service broker is essential.

One of the key actions of a service broker is \textit{service matchmaking}, that is, returning one or more suitable cloud service offers from the \textit{service providers} which fulfil the requirements of a \textit{service requester}. Cloud service offers are any computing resources which are provided on application, system and infrastructure levels on the cloud; respectively SaaS, PaaS and IaaS. These vary in functionality, quality metrics, and other non-functional properties such as legal aspects. Moreover, the exact same application deployed on different cloud infrastructures are only different in non-functional aspects. In this setting, service matchmaking is a complex problem for which the scope of matching, the detail level of results, and the service descriptions vary. The distinguishing aspects of the services can be formally described in \textit{service descriptions} using \textit{service description languages} such as Linked-USDL\footnote{Linked Unified Service Description Language} \cite{linkedUSDLSLA}, SMI\footnote{Service Measurement Index} \cite{smi2014v2}, OWL-S\footnote{Semantic Markup for Web Services} \cite{martin2004owl} and SDL-NG\footnote{Next Generation Service Description Language Framework} \cite{slawikdomain} in ongoing research projects. 

We apply the \textit{Information Systems Research Framework} \cite{hevner2004design} as our research methodology. Our work primarily targets cloud service marketplaces context. Therefore, we first evaluate existing approaches against two essential requirements of service matchmaking on service marketplaces: the ability to handle incomplete knowledge and to take the service requester's perspective into account. Constraint-based approaches presented in the papers \cite{mobedpour2013user}, \cite{kritikos2009mixed} and \cite{bacciu2010adaptive} align at best with these requirements. Following this, we evaluate the approaches which support the essential requirements with respect to the cloud service descriptions. These approaches cover the most types of properties which result from our analysis of the service descriptions in general and in our service description language\cite{slawikdomain}. Further examination reveals two improvement opportunities: list-typed quality of service (QoS) properties as well as explicit handling of preferences for lower or higher property values.

The goal of this paper is to address these issues by explicitly handling the preferences for QoS parameters and by adding support for list typed QoS parameters. Our solution builds on the idea of using constraint programming to solve the service matchmaking problem. Therefore, we present constraint models and a prototype implementation using constraint solvers which also allows fuzzy service requests.

This paper is organised as follows: The next section describes the service matchmaking problem, the analysis of the types of properties in service descriptions, and evaluates existing approaches against the requirements. Section 3 presents the constraint models, Section 4 presents their implementation using the Java Constraint Programming API (JSR-331)\footnote{Java Specification Request 331}. Section 5 evaluates our approach and presents next steps.

\section{Problem Definition}\label{sec:prob}

The requirements for service matchmakers can be defined on two levels: 1. the process as a whole, 2. the core matchmaking functionality. 

The requirements for the process depend on the application context. The application context can be cloud service marketplaces, automated service composition and inter-cloud. Service marketplaces position the service matchmaker as an assistant to a service requester who is a person. Therefore, the service matchmakers in this group build up the request step by step, consider the priorities of the service requester and categorise the results as very good, good and satisfactory (\emph{R1: Service Requester Priorities} and \emph{R2: Comprehensive Results with Matching Degrees}).


The automatic service composition context requires that the optimal service is found without user interaction and that the over-constrained requests are automatically adjusted till a service description matches the service request. For the inter-cloud context, several ways to handle the application brokering are discussed: SLA-based, trigger-action and directly managed \cite{grozev2014inter}. In the directly managed fashion, the service requester handles the deployment on multiple clouds and therefore no broker is involved. In the first two ways, a broker is involved but the service requester does not take the final decision. We argue that in a practical context the service requesters have to take the final decisions with assistance of the broker similar to service marketplaces, as they are most often liable for it - both legally and economically.


The core matchmaking functionality can be examined from three aspects: the scope of matching, incomplete knowledge and fuzziness, and the types of properties identified in the service descriptions. 

The scope of matching covers non-functional and functional properties. Functional properties matching is implemented as pre- and post-condition matching and/or API matching, which rather targets the automated service composition context and software developers. The non-functional requirements of the service requester such as interoperability, quality metrics, and legal aspects are handled as QoS matching (\emph{R3:QoS Matching}).

Service matchmakers must deal with missing values, since the service requester might not be sure about all the QoS constraints and the service provider might not supply information for all the QoS properties of the service (\emph{R4: Incomplete Knowledge}). Moreover, the service requesters should be able to define their priorities and fuzzy values with \textit{variational scope} for the constraints (\emph{R5: Fuzziness}). 

Variational scope \cite{platenius2013survey} can be introduced to a service matching approach in three ways: 
(i)The matching of service description parameters to requirements with a certain amount of tolerance---assuming most requesters would accept a service with a value slightly different than the specified value, (ii) The parameters are specified with fuzzy terms such as "good" or "very fast", (iii)The requirements are specified with their level of importance to the 
requester with qualifiers such as mandatory and optional. 

\subsection{Analysis of QoS Properties in Service Descriptions}\label{sec:typeqos}
\begin{table}[htp] 
 \begin{minipage}{\textheight} 
 \begin{tabular} 
{p{15mm}p{14mm}p{14mm}p{14mm}p{12mm}}  \toprule 
\emph{QoS Property} & \emph{App. X aaS by Provider \#1} &\emph{App. X aaS by Provider \#2}&\emph{App. X aaS by Provider \#3}&\emph{Service \newline Request}\\\midrule 
Version &5.5 & 5.6 & 5.6& $=5.6?$\\
Response time  & $<120ms$ & $<200ms$ & $<400ms$ & $<300ms$ \\ 
Storage in Free Version & 0GB& 15GB & 20GB & $>5GB$\\ 
Availability & $>99.99\%$& $>99.95\%$ & $>99.95\%$& $>99\%$ \\
Establishment Year &2010 & 2005 & 2012& $>2009$\\
Pricing &per dyno-hour & per number of requests & per hour & per hour \\
Compatible Browsers & Explorer, Chrome, Firefox & Explorer, Chrome, Firefox, Safari & Explorer, Safari & Explorer, Firefox, Safari \\ \bottomrule 
 \hline \end{tabular} \end{minipage} 
 \caption{An Example Service Matchmaking Problem} \label{tab:example} 
 \end{table} 

This paper analyses the service description 
languages/non-functional properties frameworks and parameters commonly used in SLA description languages, especially SMI \cite{siegel2012cloud} and 
CRF \footnote{Cloud Requirements Framework}\cite{repschlaeger2012cloud} and the SDL-NG\cite{slawikdomain} developed as part of the project TRESOR\footnote{Trusted Ecosystem for Standardized and Open cloud-based Resources}\cite{tresor2014}.

Table \ref{tab:example} shows examples for different types of properties in three service descriptions and a service request. The example describes the non-functional properties of a database application deployed on different cloud infrastructures by different cloud providers. The service request comprise the constraints on service properties on the right most column of Table \ref{tab:example}. 

For \textit{version}, the specification must be equal to the service request which is a numeric value, we name these \textit{discrete numeric value} and \textit{discrete value matching}. For \textit{response time}, the service description guarantees an upper limit. It can be assumed that no discrete value matching will be performed and the lower limit of the request can be ignored. We will refer to these as 
\textit{low-value preferred properties}. For \textit{storage in free version}, higher values are preferred, only interval matching with a lower limit is needed and the service description guarantees a lower limit. Similarly, the upper limit of the request can be ignored. We name these \textit{high-value preferred properties}. For the low-value preferred and high-value preferred properties \textit{interval matching} is applied with the assumptions stated above. In addition, for some properties some service requesters prefer higher values and some lower values as in the case of \textit{establishment year}. We will call this \textit{requester defined preference}. The requester defines an upper or lower limit, and it will be matched to the service  value or range e.g. $51<x<200$ or 100+. An example for this is the number of employees in a service provider profile. \textit{Pricing} is an example for an enumeration, the service 
specification can only take one of the values in the predefined list. In the case of feature lists both the service request and the service specification can take multiple values from the list as given in the
\textit{compatible browsers} example.

Based on the types of properties analysed above, we define the subproblems of service matching as follows: discrete value matching, feature list matching, interval matching, and discrete value matching with soft constraints (\textit{R6: QoS Matching Data Types Coverage}). 

\section{Related Work}
Some approaches do not take incomplete knowledge (R1) into account in the service descriptions and service requests \cite{yu2007efficient}, \cite{sarang2012clustering}. D'Mello et al. present an approach \cite{d2008semantic} which compares the services with each other without requester's constraints.

Table \ref{tab:treReqTable} examines QoS Matching data types coverage in related work. Although some subproblems are identified and addressed, existing approaches have shortcomings. All three approaches suggested in  \cite{kritikos2008evaluation}, \cite{mukhija2007qos} and \cite{mobedpour2013user} assume that the properties are either low-value or high-value preferred, although there are properties for which the service requester might be searching for exact values. The approach presented by Kritikos et al. \cite{kritikos2008evaluation}, \cite{kritikos2009mixed} improve the algorithm presented by Ruiz et al. \cite{ruiz2005improving} with the advanced categorisation of results. Moreover, they do not consider enumerations and fuzziness. None of them support feature lists leaving this subproblem out. 


\begin{table}[htp] 
\begin{minipage}{6cm}
\begin{tabular} 
{p{16mm}p{9mm}p{5mm}p{5mm}p{7mm}p{5mm}p{10mm}}  \toprule 
\emph{Research Work} & \emph{Features} &&&&& Maturity Level \\\midrule 
Matchmaker    &  Discrete numeric data & Enumer- ation & Intervals & Fuzziness & Feature Lists & \\ 
\cite{mobedpour2013user}  & yes & no & yes & yes& no& prototype \\ 
\cite{bacciu2010adaptive}, \cite{mukhija2007qos} & yes & yes & yes & yes \footnote{false negatives for super matches}  & no & prototype and theory \footnote{fuzzy not in prototype} \\
\cite{kritikos2008evaluation},\cite{kritikos2009mixed}, \cite{kritikos2007semantic}& yes & no & yes & no &no&prototype\\
\cite{wang2009qos} &no & no & no& yes&no&theory \\
 \bottomrule 
 \end{tabular} 
 \end{minipage}
 \caption{Data Types of Properties and Their Handling in Related Work} \label{tab:treReqTable} 
 \end{table} 
The interval matching approach suggested in \cite{mukhija2007qos} determines if the property in question is high-value preferred or low-value preferred based on the values that the service requester specified: If the most preferred value of the service requester is smaller than the least 
preferred value, than the property is assumed to be low-value preferred. Firstly, this sets the prerequisite that the service requester knows which values are better. However, this might not be the case. Secondly, the service requester specifies both an upper limit and a lower limit in all cases. If the property is low-value preferred, the service requester should not be prompted to specify a lower limit. Besides, the BV calculation would not work if the service requester specifies the same value for the most preferred and the least preferred values.

The trapezoidal fuzzy numbers approach \cite{bacciu2010adaptive} would deliver faulty results for low-value preferred properties and high-value preferred properties. However, it can be applied to the cases where the properties do not have broadly accepted tendencies.

\section{Constraint Models}\label{sec:cpModels}
This section presents the constraint models which address the subproblems defined in Section \ref{sec:typeqos}.
Bockmayr and Hooker \cite{cpElement} define element constraints as follows: 

"$element(i, l, v)$ expresses that the $i-th$ variable in a
list of variables $l = [x 1 , . . . , x n ]$ takes the value $v, i.e., x i = v.$" 

We use element constraints to model the service matchmaking problem.

\subsection{Model \#1} 
\begin{figure*}[htpb]
\centering
\captionsetup{justification=centering,margin=2cm}
 \includegraphics[width=17cm]{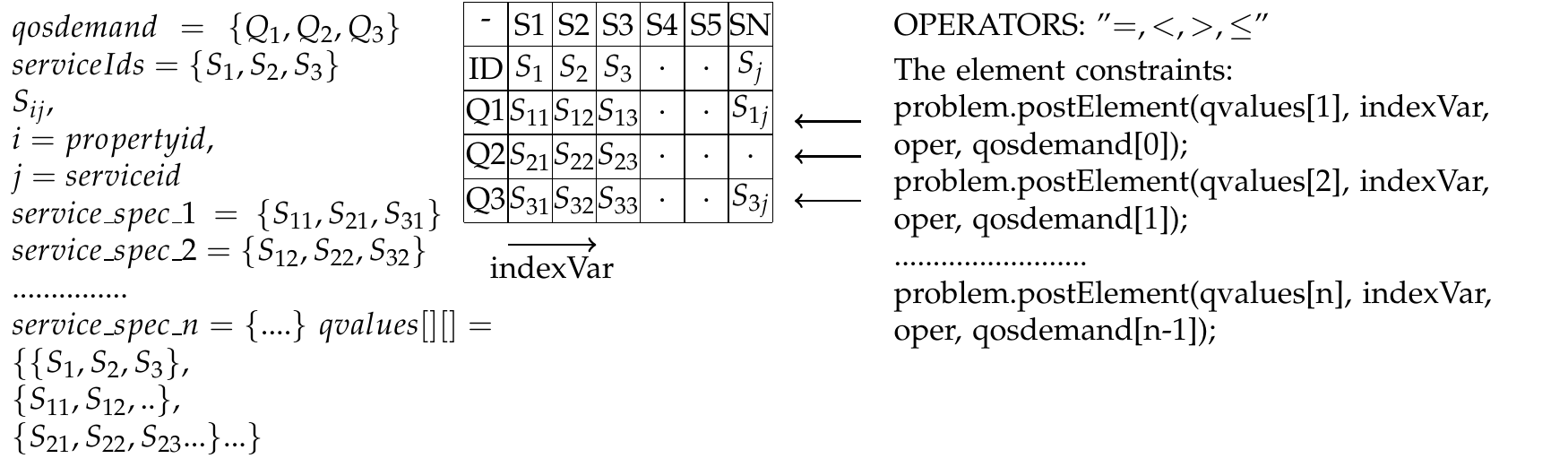}
  \caption{Constraint Solving Problem Model as a Matrix}
  \label{fig:matrix1}
\end{figure*}

Fig. \ref{fig:matrix1} illustrates the first constraint model developed in this paper. Each row in the matrix contains the values of 
service specifications for the QoS property which is on the left most column.
For each row, an element constraint is defined which adds the condition:
\begin{flalign}
qvalues[i]"operator"qosdemand[i-1]
\end{flalign} 
to the constraint solving problem. The operators can be adjusted from the list of available operators according to the specific purpose of the CSP 
utilising the model. The first row in the qvalues matrix is the array of service ids. For this reason, the element constraints begin with the 
second row. Note that the QoS request and service specifications are modelled as Java integer arrays, but not as variables, since the values for 
those are fixed and the variable the model sets as unknown is the index variable.

\subsection{Model \#2}
The second constraint model takes another perspective to the service matching problem. Its main difference to
the first model is that it takes properties as JSR-331 variables whose domain is an array consisting of the values from the service specifications. This means each row in the matrix is defined as a variable:

\begin{lstlisting}[frame=single,language=Java,caption=Discrete Value Matching Model]  
problem.variable(”property1”,qvalues[1]);
problem.variable(”property2”,qvalues[2]);
\end{lstlisting}
 
 The resulting CSP searches for appropriate values of the property variables and the index variable. To ensure the integrity of a service description, 
additional constraints are needed, since a Java array cannot get a JSR-331 variable as an index and the index of a Java array cannot be tracked by the
JSR-331. These constraints state that if serviceId has a certain value, the property value can have only one value from its domain, which is the 
value in domain[serviceId].

\noindent\begin{minipage}{.45\textwidth}
\begin{equation}\label{eq:1}
\begin{aligned}
p:serviceId=x,\\
q:property1=domain[x]\\
p \equiv q
\end{aligned}   
\end{equation}
\end{minipage}\hfill%

The relation can be expressed with logical equation since $p$ and $q$ are either both true or both false to achieve a result true.
With the element constraints, it was possible to do this without additional constraints, however element constraints are only available as hard constraints. 

\section{Implementation}\label{sec:impl}
The prototype implementation uses JSR-331\cite{jsr331} with Choco Solver \cite{Choco} to implement the constraint models. Discrete value matching with hard constraints, interval matching for negative and positive
tendencies and feature list matching is realised using Model \#1. Discrete value matching with soft constraints
is realised using Model \#2. The models are described in Section \ref{sec:cpModels}. The implementation source can be found in our repository. \footnote{\url{https://github.com/TU-Berlin-SNET/cloud-service-matcher}}

Our implementation employs four methods which model the problem differently: \textit{buildModel} for exact matching with only hard constraints,
\textit{buildModelSoftAsBool} for matching with soft Boolean constraints, \textit{buildModelSoftDifference} for matching with soft constraints according to the difference
between values of the service offer and request, and buildSimpleModelDifference which is an enhanced version of buildModelSoftDifference. 

\subsection{Discrete Value Matching with Hard Constraints}\label{subs:discH}
For discrete value matching with hard constraints, the implementation makes use of Model \#1 which is described in Section \ref{sec:cpModels}. Each row in the matrix contains the discrete numeric values of service specifications. For each row there is an element constraint which adds the condition qvalues[i]=qosdemand[i-1] to the CSP.
For example, indexVar=1 is in the solution set since qvalues[1][1]=2 equals to qosdemand[0].

\begin{lstlisting}[frame=single,language=Java,caption=Discrete Value Matching Model]  
Var indexVar = matching
.variable("serviceIndex", 0, serviceIndexMax);
for (int j = 1; j < qosdemand.length; j++) {
	matching.postElement(qvalues[j], 
	indexVar, "=", qosdemand[j - 1]);
	}
\end{lstlisting}

For interval matching only the operator has to be changed: for properties with positive tendency $>=$ and for negative tendency $<=$.

\subsection{Discrete Value Matching with Soft Constraints}\label{sec:discreteSoft}
\textit{buildModelSoftAsBool} introduces fuzziness with the third option of variational scopes described in Section \ref{sec:prob}. It creates the negation of the element constraints defined in \textit{buildModel} described in Section \ref{subs:discH}. In contrast to hard constraints,
these are not posted, instead an optimisation objective is defined using them. If the constraint is satisfied, the constraint method returns 1, if not 0. If the service specification value is not equal to the service request value the value 1 is then multiplied by the weight for the QoS parameter that the service requester specified.
The violation is calculated for each element constraint and then added to the violation sum. 
The solver returns the service index with the minimum violation sum which is the optimisation objective.  

The condition checks if the service specification value exactly matches the requirement value and if not adds up to the violation sum. In some cases, this is not enough since the requester might specify an approximate value for a requirement and the results would be still fulfilling even if they are slightly different
than the requirement which is described as fuzziness with variational scope's first option above. To provide this kind of fuzziness, the optimisation objective must be the difference between the service specification value and the requirement value. 

In this case, the optimisation objective can be defined as: 
  
\begin{equation}
\begin{aligned}\label{eq:2} 
i=qoSPropertyId\\
j=serviceId\\
|S_{ij}-Q_{i}|
\end{aligned}   
\end{equation}

The coding experiments in \textit{buildModelSoftDifference} with element constraints as soft constraints showed that if the element constraints are not posted, then the variable \textit{serviceIndex} is not constrained, so they were not 
effective. For this reason, the element constraints were removed from the problem and linear constraints were added to ensure service id and service value 
bindings. In other words, for defining the 
optimisation as the difference and getting consistent results, the Model \#2 was designed (see Section \ref{sec:cpModels}). 

Getting only one solution is not suitable for the service matching problem, since it does not provide all, if there are equally optimal solutions.
As a workaround, the service matcher uses the CP solver to find all the solutions as a list and 
orders them according to their values for violation from minimum to maximum. This way, it can be seen if there are some solutions with the same violation value 
and appropriately evaluated. 

\subsection{Feature List Matching}\label{sec:featureLM}
For feature list type of constraints, a constraint solving problem per constraint must be defined.

This time, the index variable shows the elements where in the feature list QoS specification the values match with the feature list QoS constraint. We create and post a new element constraint and the default solution logger lists the values for \textit{indexVar} and \textit{var} which satisfy the constraint. 

The matching degree is calculated based on the size of the solution set and further explained in Section \ref{ssubs:rankingF}. This implementation handles all the items in the required list equally. 
\begin{lstlisting}[frame=single,language=Java, caption= Feature List Matching Example\label{lst:featureExample} ]
String[] browsers = { "explorer", "firefox", "chrome",
 "safari", "opera" };
int[] providedBrowsers = { 1, 2, 0 };
int[] requiredBrowsers = { 0, 2, 3 };
INFO:
providedIndex[1] query[2]
providedIndex[2] query[0]
matching provided browser value:0 name:explorer
matching provided browser value:2 name:chrome
\end{lstlisting}

Listing \ref{lst:featureExample} shows that the solution set has two elements: $providedBrowsers[1]=2$ and $query=2$, $providedBrowsers[2]=0$ and $query=0$. The matching degree
is calculated based on the size of the solution set and further explained in Section \ref{ssubs:rankingF}.


\subsubsection{Ranking for Feature List Matching}\label{ssubs:rankingF}
\paragraph*{Matching Degree}
The Feature List Constraint contains the number codes for a list of required items. Accordingly, the Feature List QoS Specification contains the list of number
codes that the service offers for that property. 
$P$ is the set of provided browsers. $R$ is the set of requested browsers. $S$ is the set of solutions, and can be described as the intersection of provided and requested sets.

\begin{flalign}
S=P \cap R  \nonumber \\
P=\emptyset \implies NOSPEC \nonumber \\
S=\emptyset \implies FAIL \nonumber \\
\left\vert{R}\right\vert >  \left\vert{S}\right\vert \implies PARTIAL  \\
\left\vert{R}\right\vert =  \left\vert{S}\right\vert, \left\vert{P}\right\vert = \left\vert{S}\right\vert \implies EXACT \nonumber \\
\left\vert{R}\right\vert =  \left\vert{S}\right\vert, \left\vert{P}\right\vert > \left\vert{S}\right\vert \implies SUPER \nonumber
\end{flalign}

An example for this type of QoS property is the list of compatible browsers. 
\paragraph*{Ranking Rules}
The ranking rules define how many points a service description gets according to the matching degree of its QoS specification. These are defined in the \textit{Evaluator} classes.
For example, an \textit{ExactEvaluator} gives 2 points to the QoS specification. These rules can be changed at the corresponding Evaluator without touching other
parts of the code. Table \ref{tab:matchingdegreeF} shows the scheme for the ranking rules. At the time of writing, soft constraints calculate the violation based on the 
weights and the ranking for the hard constraints add scores for each matching QoS specification independently. 

The final score of a service can be calculated as
$(sum Of Scores) - (sum Of Violations)$ as also shown below.
\begin{flalign}
\sum_{i=1}^{j} s_i - \sum_{i=1}^{k} v_i \nonumber \\
j=\mbox{number of hard QoS constraints} \nonumber \\
k=\mbox{number of soft QoS constraints} \\
s_i= \mbox{score for the QoS Specification} \nonumber \\
v_i=\mbox{violation for the QoS Specification} \nonumber
\end{flalign}

\begin{table}[htpb]
\begin{tabular}{llllll}
\toprule
Provided & Requested & Solutions & Matching Degree & Ranking Rules \\
\midrule
0,1,4    & 0,1       & 0,1       & SUPER           &  3 points\\  
0,1      & 0,1       & 0,1       & EXACT           &  2 points\\
0,4      & 0,1       & 0         & PARTIAL         &  1 point\\
2,3      & 0,1       & none      & FAIL            &  0 points\\
none     & 0,1       & none      & NOSPEC          &  0 points\\ \bottomrule
\end{tabular}
\caption{Matching Degree Examples and Ranking Rules}\label{tab:matchingdegreeF}
\end{table}
\vspace*{-\baselineskip}
\section{Evaluation}
This section describes the goal-free comparison of our approach with other processes. 
\begin{table} 
    \begin{tabular}{p{35mm}p{10mm}p{30mm}}               \toprule 
    Description & Design \newline Artifact& Evaluation Method  \\\midrule
    process of matching with its inputs and outputs                 & method           & goal-free comparison with other processes  \cite{scriven1991evaluation}                  \\
    our implementation of the process                  & instantiation          & testing                  \\
    \bottomrule
    \end{tabular}
    \caption{Evaluation Methods}
 \label{tab:evalMethodsTable}
\end{table}


Our solution is designed analysing QoS properties in service descriptions, therefore it addresses \textit{R3:QoS Matching}. Moreover, it supports the most frequent combinations of the data types in its core matching functionality addressing all subproblems in Table \ref{tab:treReqTable}.
We address \textit{R4: Incomplete Knowledge} and \textit{R5:Fuzziness} by allowing service requesters to define both hard and soft constraints. We provide two types of soft constraints: (i)the equality of discrete values in the specification and the request as Boolean with weights also addressing \textit{R1: Service Requester Priorities}, (ii)the distance of the value in the specification to the specified value.
\textit{R2: Comprehensive Results with Matching Degrees} is addressed by the service matcher since it includes all evaluations of service specifications within the service descriptions which is easily accessible if needed. The implementation for the exact matching of two intervals is left for future
work, since the priority was feature lists due to the TRESOR project context.

This paper contributes to the constraint-based service matching methods suggested in \cite{kritikos2009mixed}, \cite{mobedpour2013user} and the approaches developed in the Dino project \cite{mukhija2007qos},\cite{bacciu2010adaptive} by developing a better picture of the properties to be matched and diagnosing various assumptions made in the definitions of the preferences of the service requesters. It supports the view that models the service matching problem as an optimisation problem and that the use of 
constraint solvers is especially suitable for the implementation of soft constraints. It challenges the views which implicitly assume that the QoS properties are either low-value preferred or high-value preferred.

\section{Conclusion and Future Work}
In this paper, we analyse the requirements for service matchmaking approaches on the process and on the core functionality levels. We identify that the application context has a defining effect on how the service requester interacts with the system and how the results are further categorised. This serves as a tool to evaluate each service matchmaker in its context. On the core functionality level, we analyse the QoS properties and identify the subproblems discrete value matching, feature list matching, interval matching, and discrete value matching with soft constraints. Our prototype implementation provides solutions for all these subproblems. We suggest that the low-value, high-value, and neutral preferences for QoS properties are explicitly stated when documenting the target properties for matchmaking functionalities.

As future work, a case study might be useful to identify additional requirements in a practical context. 
Moreover, we will look into the specific requirements of intercloud application brokering and extend the service matchmaker accordingly.
\section*{Acknowledgment}
This work is supported by the Horizon 2020 EU funded Integrated project
CYCLONE\footnote{cyclone-project.eu}, grant number 644925.

\bibliography{main}{}
\bibliographystyle{IEEEtran}


\end{document}